\documentclass[preprint,showpacs,showkeys,preprintnumbers,amsmath,amssymb]{revtex4}
\usepackage[dvips]{graphicx}
\usepackage{dcolumn}
\usepackage{bm}

\newcommand{\astar}{\ensuremath{a_*}}

\begin{document}

\preprint{}

\title{Rotating black holes at future colliders II:\\
Anisotropic scalar field emission}

\author{Daisuke Ida}
\email{daisuke.ida@gakushuin.ac.jp}
\affiliation{Department of Physics, Gakushuin University, Tokyo 171-8588, Japan}
\author{Kin-ya Oda}
\email{odakin@th.physik.uni-bonn.de}
\affiliation{Physikalisches Institut der Universit\"at Bonn, Nussallee 12, Bonn 53115, Germany}
\author{Seong Chan Park}%
\email{spark@mail.lns.cornell.edu} \affiliation{Institute for High
Energy Phenomenology, Floyd R. Newman Laboratory for Elementary
Particle Physics, Cornell University, Ithaca, N.Y. 14853, U.S.A.}

\date{\today}

\begin{abstract}
This is the sequel to the first paper of the series,
   where we have discussed the Hawking radiation from five-dimensional rotating
   black holes for spin $0$, $1/2$ and $1$ brane fields
   in the low frequency regime.
   We consider the emission of a brane localized scalar field from rotating black holes
   in general space-time dimensions
   without relying on the low frequency expansions.
\end{abstract}

\maketitle

\section{Introduction}
Black hole is one of the most
important key objects in theoretical physics.
Though its quantum behavior and
thermodynamic property have played great roles
in the path to understand yet unknown quantum theory of gravity
(see e.g.\ Refs.~\cite{David:2002wn,Padmanabhan:2003gd,Veneziano:2004er}),
a direct experimental
test had been believed almost impossible.
Recently, the scenarios of large~\cite{Hamed:1998rs}
and warped~\cite{Randall:1999ee} extra dimension(s)
have led to an amazing possibility of producing black holes
at future colliders
with distinct signals~\cite{Giddings:2001bu,Dimopoulos:2001hw}
     (see also Refs.~\cite{Argyres:1998qn,Banks:1999gd}
     for studies before the observation~\cite{Emparan:2000rs}
     that black holes radiate mainly on the brane).

When the center-of-mass (CM) energy of a collision exceeds the
Planck scale, which is of the order of TeV here, the cross section
is dominated by a black hole production~\cite{'tHooft:1987rb},
which is predicted to be of the order of the geometrical
one~\cite{Hsu:2002bd,Eardley:2002re,Yoshino:2002tx,Giddings:2004xy},
increasing with the CM energy. In this trans-Planckian energy
domain, the larger the CM energy is, the larger the mass of the
resulting black hole is, and hence the better its decay is treated
semi-classically via Hawking radiation~\cite{Hawking:1974sw}. Main
purpose of this series of work is to discuss such decay signals in
the hope that these will serve as the basis to pursue stringy or
quantum gravitational corrections to them.

In previous publications, we have pointed out that the production
cross section of a black hole increases with its angular momentum,
so that the produced black holes are highly
rotating~\cite{Ida:2002ez,Ida:2005zi} (see also
Ref.~\cite{Park:2001xc} for an earlier attempt). The form factor
for the production cross section, taking this rotation into
account~\cite{Ida:2002ez}, is larger than unity and increases with
the number of dimensions $D=4+n$. The result is in good agreement
with an independent numerical simulation of a classical
gravitational collision of two massless point
particles~\cite{Yoshino:2002tx}. We note that this form factor is
hardly interpretative without considering the angular momentum. It
is indispensable to take into account the angular momentum of the
black hole when we perform a realistic calculation of its
production and evaporation.

Black holes radiate mainly into the standard model fields that are
localized on the brane~\cite{Emparan:2000rs}. In the previous
paper~\cite{Ida:2002ez}, we have shown that the massless brane
field equations with spin zero, one-half and one, i.e. for all the
standard model fields, are of variables separable type in the
rotating black hole background. Then we obtained the analytic
expressions for the greybody factor of $D=5$ (Randall-Sundrum 1)
black hole by solving the master equation under the low energy
approximation of radiating field. In this work we present a
generalized result to higher dimensional black hole in $D\geq 5$
for brane localized scalar field without relying on the low energy
approximation and discuss its physical implications.

This paper is the longer version of the brief report~\cite{Ida:2005zi}.
We also note that the related works by Harris~\cite{Harris:2004mf}
and by Harris and Kanti~\cite{Harris:2005jx} appeared more recently.

\section{Brane scalar field emission} 
A brane-localized scalar field $\Phi$ in the higher $(4+n)$-dimensional rotating
black hole background~\cite{Myers:1986un}
can be decomposed into the radial and angular parts
$R(r)$ and $S_{\ell m}(\vartheta)$, respectively~\cite{Ida:2002ez}
\begin{eqnarray}
\Phi=R(r)S_{\ell m}(\vartheta)e^{-i\omega t+im\varphi},
\end{eqnarray}
where the Boyer-Lindquist coordinate
$(t,r,\vartheta,\varphi)$ reduces to the spherical coordinate at
spatial infinity
and $\ell,m$ are the angular quantum numbers.
The resultant equations are shown to be separable~\cite{Ida:2002ez}.
The angular part $S_{\ell m}$ obeys the equation for the spheroidal harmonics
while the radial equation becomes
\begin{align}
  \left[{d\over dr}\Delta{d\over dr}+
    {[(r^2+a^2)\omega-ma]^2\over\Delta}
    +2ma\omega-a^2\omega^2-A
    \right]R
  &= 0.
  \label{eq:Teukolsky}
  \end{align}
where
\begin{align}
  \Delta(r) &= (r^2+a^2)-(r_h^2+a^2)\left({r\over r_h}\right)^{1-n},
  \end{align}
with $r_h$ and $a$ being the horizon radius and the rotation parameter of the black hole, respectively. Note that $\Delta(r_h)=0$.

The power spectrum of the Hawking radiation is governed by, for each scalar mode,
\begin{align}
  {dE\over dt\,d\omega\,d\cos\vartheta}
    &= { \omega \Gamma_{\ell m}\over e^{(\omega-m\Omega)/T}-1}|S_{\ell m}(\vartheta)|^2,
  \end{align}
where $\Omega$ and $T$ are the angular velocity and the Hawking temperature of black hole
\begin{align}
  \Omega &= {a_*\over(1+a_*^2)r_h}, &
  T      &= {(n+1)+(n-1)a_*^2\over 4\pi(1+a_*^2)r_h},
  \end{align}
with $a_*=a/r_h$.
The $\Gamma_{\ell m}$ is the greybody factor
that determines the departure from black body spectrum,
which is the main object of this paper.
One immediate observation is that the contribution from $m>0$
modes dominates over that from $m<0$ modes
in rapidly rotating case: $\Omega\gg\omega$.

\section{Numerical evaluation of greybody factor}
The asymptotic forms of the radial wave function
at the near horizon (NH) and far field (FF) limits,
$r\rightarrow r_h$ and $r\rightarrow\infty$ respectively,
are~\cite{Ida:2002ez}
\begin{align}
  R_\text{NH} &= Y_\text{in}e^{-ikr_*} +Y_\text{out}e^{ikr_*},  \\
  R_\text{FF} &= Z_\text{in} {e^{-i\omega r_*}\over r}
                +Z_\text{out}{e^{ i\omega r_*}\over r},
                \label{FFeq}
  \end{align}
where $k=\omega-ma/(r_h^2+a^2)$ and the tortoise coordinate $r_*$ is defined
by $r_*(r)\rightarrow r$ for $r\rightarrow\infty$ and
\begin{align}
  {dr_*\over dr} &= {r^2+a^2\over\Delta(r)}.
  \end{align}
We obtain the greybody factors in the following steps.
\begin{enumerate}
    \item
    Put the purely ingoing boundary condition $Y_\text{out}=0$ by imposing
    \begin{align}
      R(r_0)  &\rightarrow e^{-ikr_*(r_0)}, &
      R'(r_0) &\rightarrow -ik{r_0^2+a^2\over\Delta(r_0)}e^{-ikr_*(r_0)},
      \end{align}
    at the NH region $r_0 = r_h(1+\epsilon)$.
    \item 
    Numerically integrate the master equation~\eqref{eq:Teukolsky}
    from the NH region to FF regime $r=r_\text{max}$.
    \item 
    Perform a least squares fit
    to the obtained data by the function \eqref{FFeq} around $r=r_\text{max}$
    to get $Z_\text{in}$ and $Z_\text{out}$.
    \item
    Finally greybody factor for the ($\ell,m$) mode is given by
    the absorption rate
    \begin{align}
      \Gamma_{\ell m}=1-\left|{Z_\text{out}\over Z_\text{in}}\right|^2.
      \end{align}
  \end{enumerate}
For the angular eigenvalue $A$, we employed the small $a\omega$ expansions
up to 6th order in \cite{Seidel:1988ue}.
(The last 6th order term in the expansion is less than a few percent
of the leading order term
    at $a\omega\lesssim 3$ for the $(\ell,m)=(0,0)$ and $(2,0)$ modes
and at $a\omega\lesssim 4$ for all the other modes.)
We have also performed above procedure in
the ingoing Kerr-Newman coordinate as a cross-check.

\section{Results}

\begin{figure}[tbp]
\begin{center}
\includegraphics[width=0.6\linewidth]{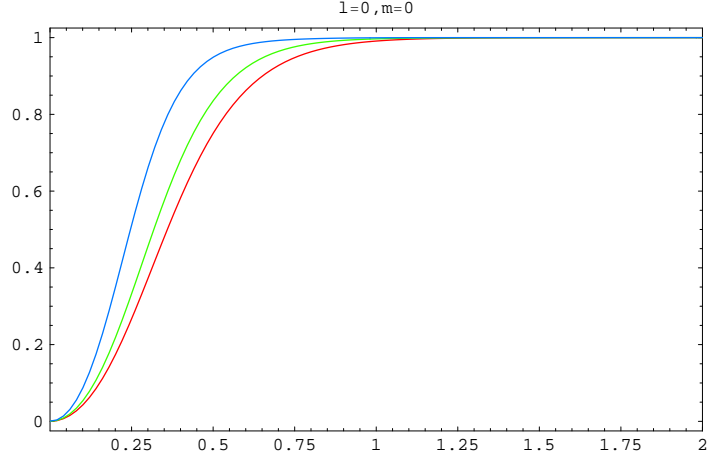}
\end{center}
\caption{Greybody factor for five dimensional black hole
for the brane scalar emission into the $\ell=0$ mode. The red, green and blue
curves correspond to $\astar=0$, $0.5$ and $1.0$, respectively.  }
\label{greybody_0}
\end{figure}
\begin{figure}[tbp]
\begin{center}
\includegraphics[width=0.4\linewidth]{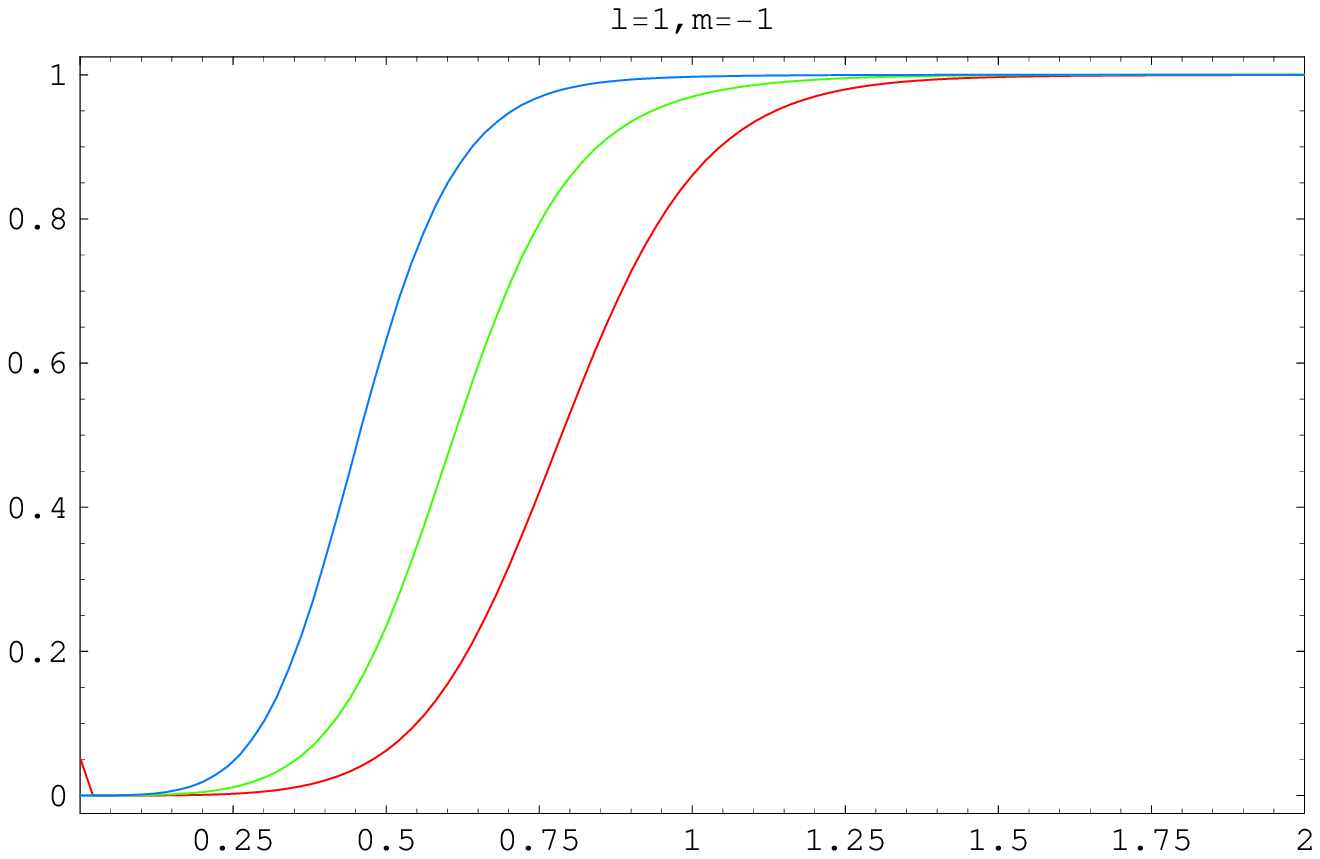}
\includegraphics[width=0.4\linewidth]{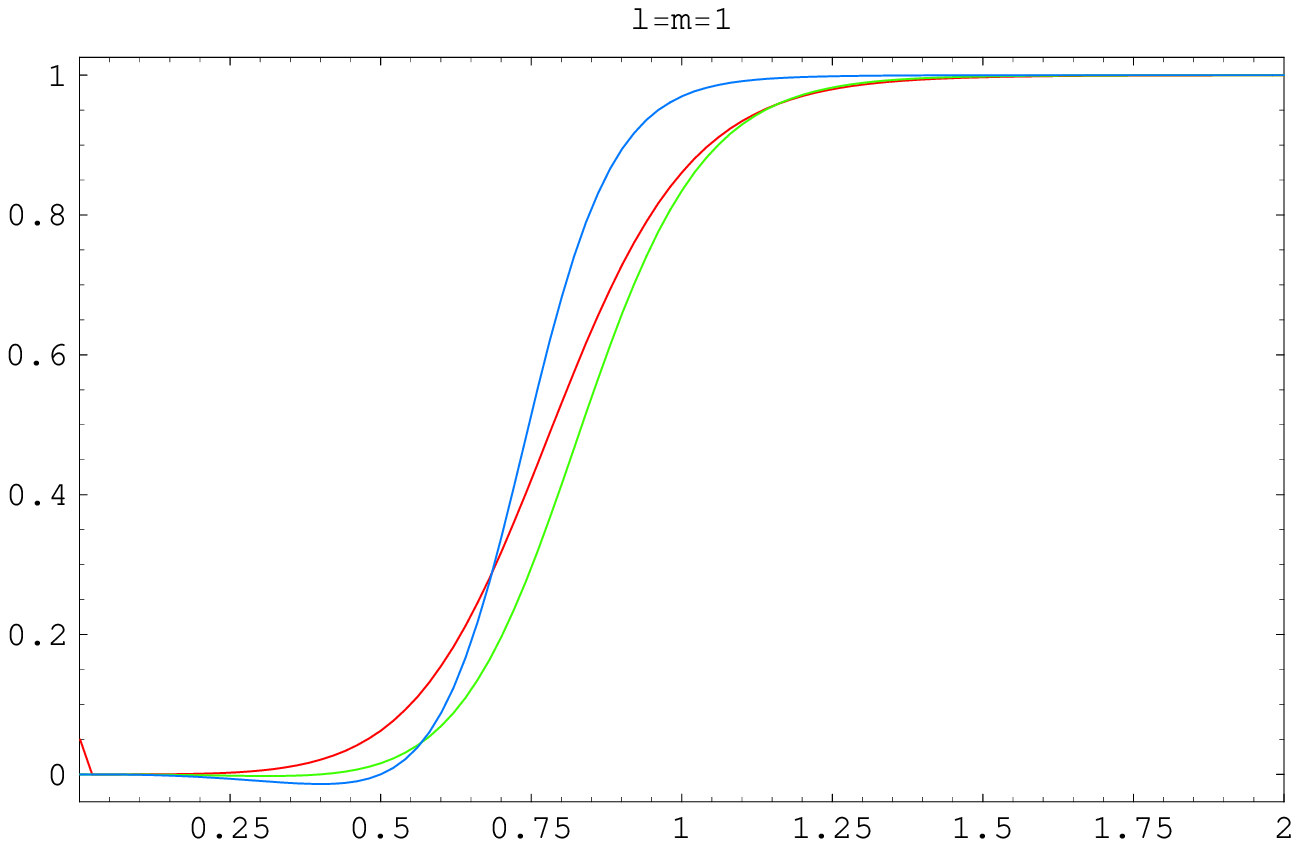}
\includegraphics[width=0.4\linewidth]{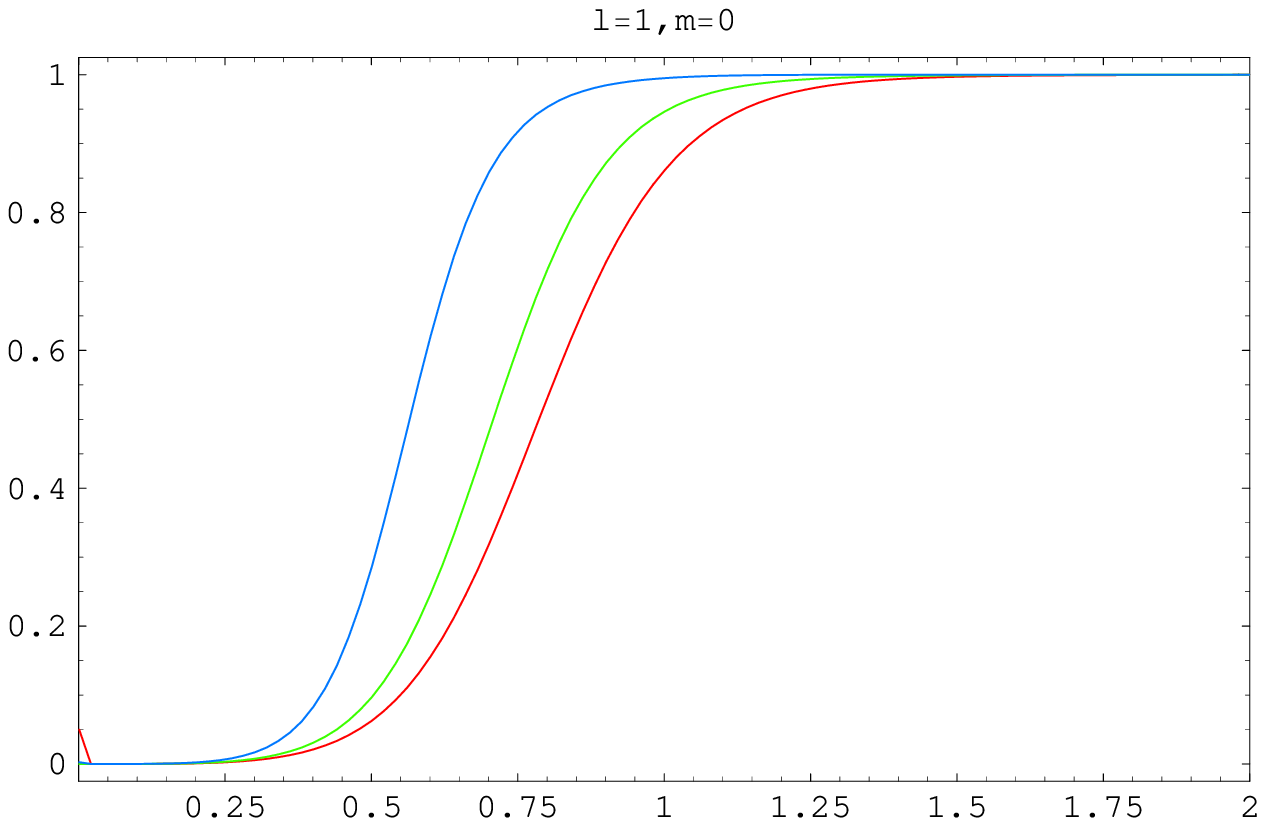}
\end{center}
\caption{Greybody factors for the $D=4+n=5$ black hole
for the brane scalar emission into the $\ell=1$ modes with $m=-1,1$ and $0$
for the upper-left, upper-right and
lower graphs, respectively.
The $m=1$ mode
shows the superradiance, namely a negative greybody factor,
in the low-frequency region $\omega<ma/(r_h^2+a^2)$.
The red, green and blue curves correspond to $\astar=0$, $0.5$ and $1.0$.}
\label{greybody_1}
\end{figure}
\begin{figure}[tbp]
\begin{center}
\includegraphics[width=0.4\linewidth]{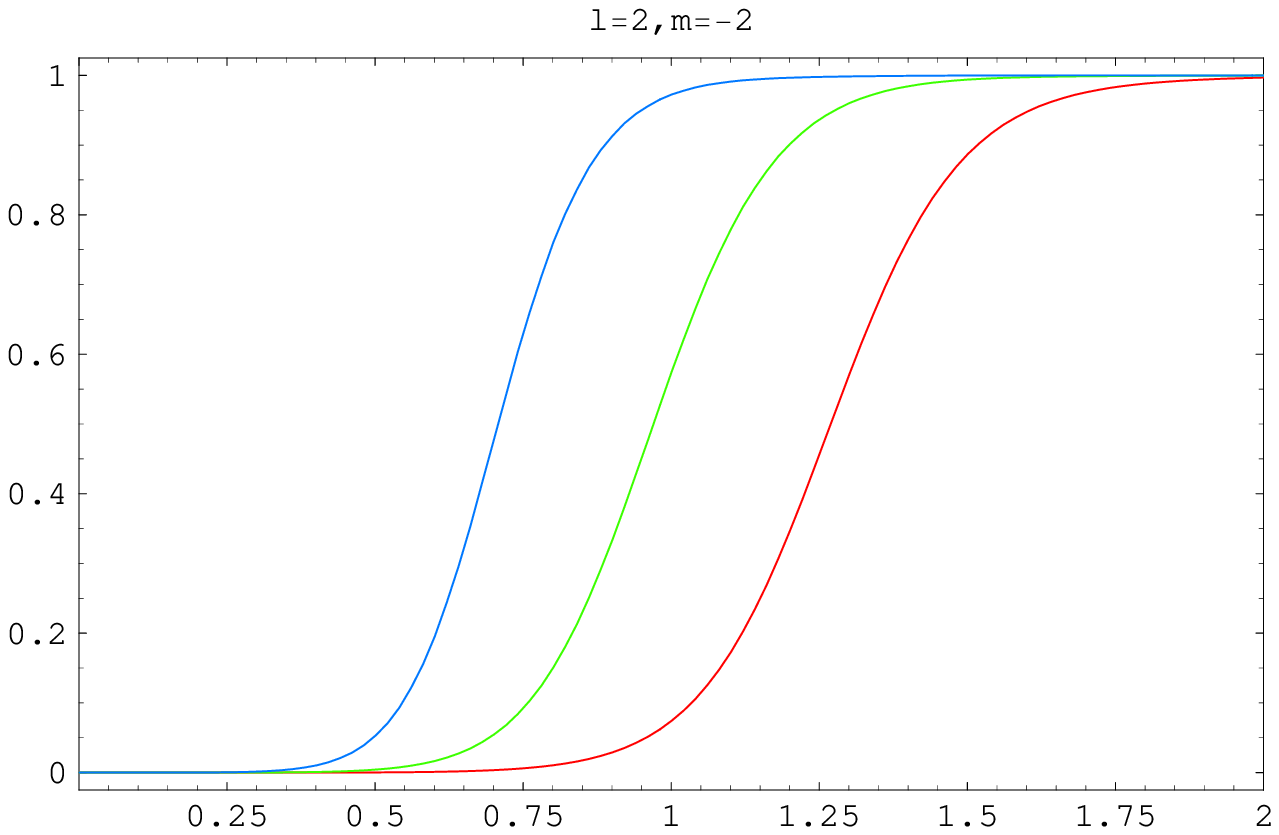}
\includegraphics[width=0.4\linewidth]{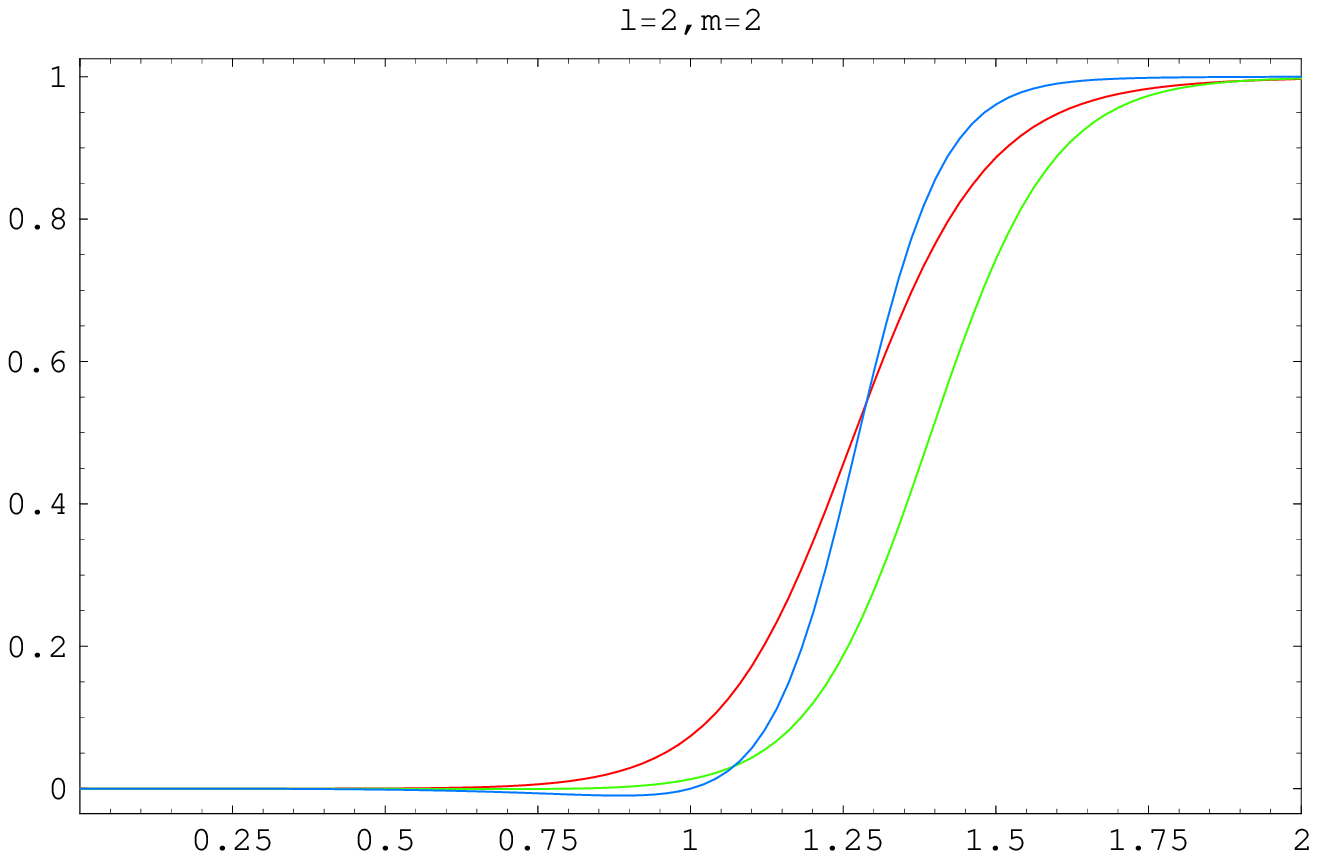}
\includegraphics[width=0.4\linewidth]{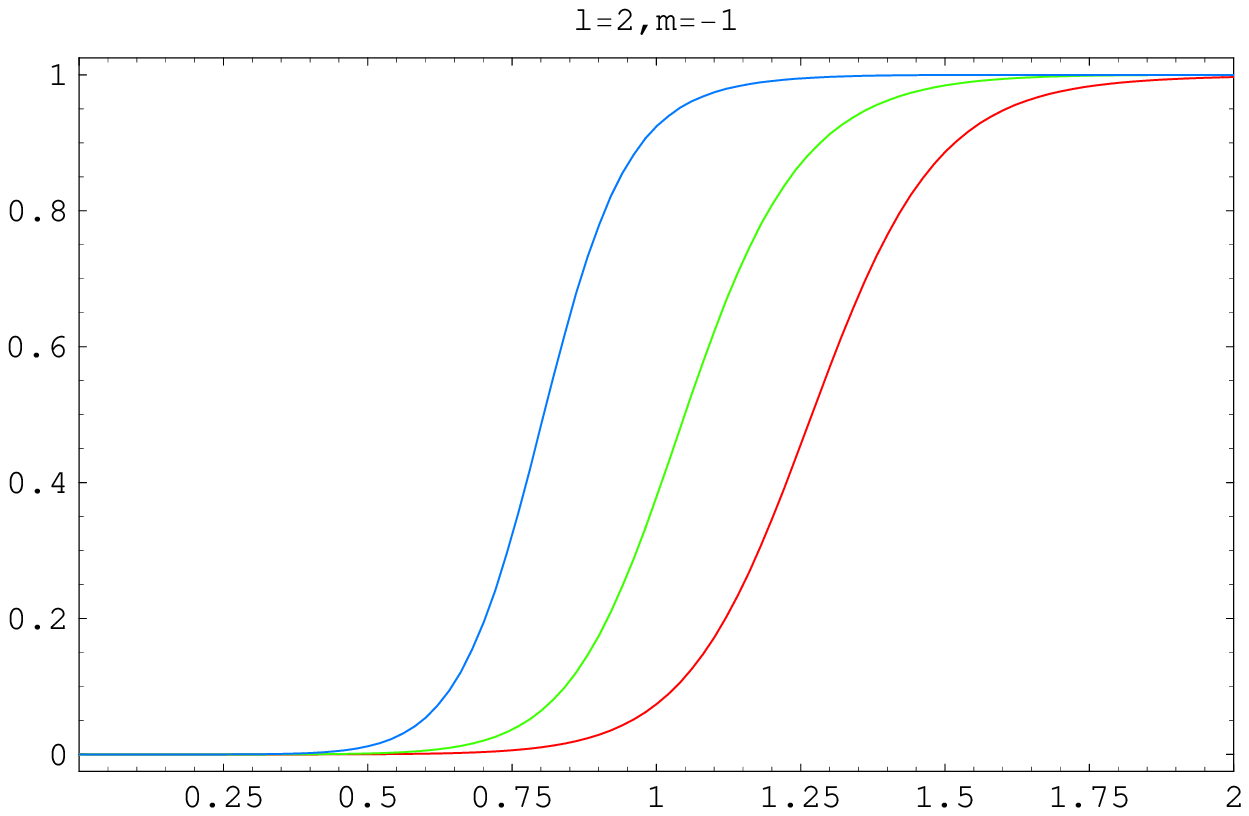}
\includegraphics[width=0.4\linewidth]{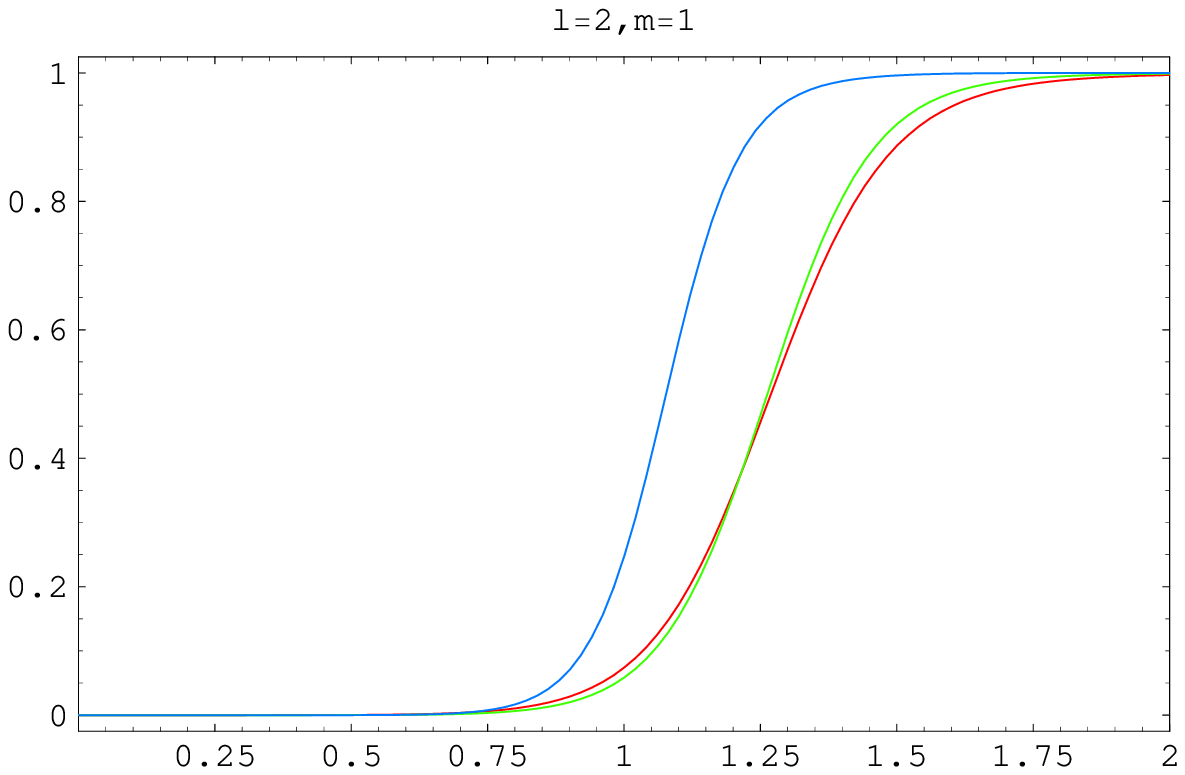}
\includegraphics[width=0.4\linewidth]{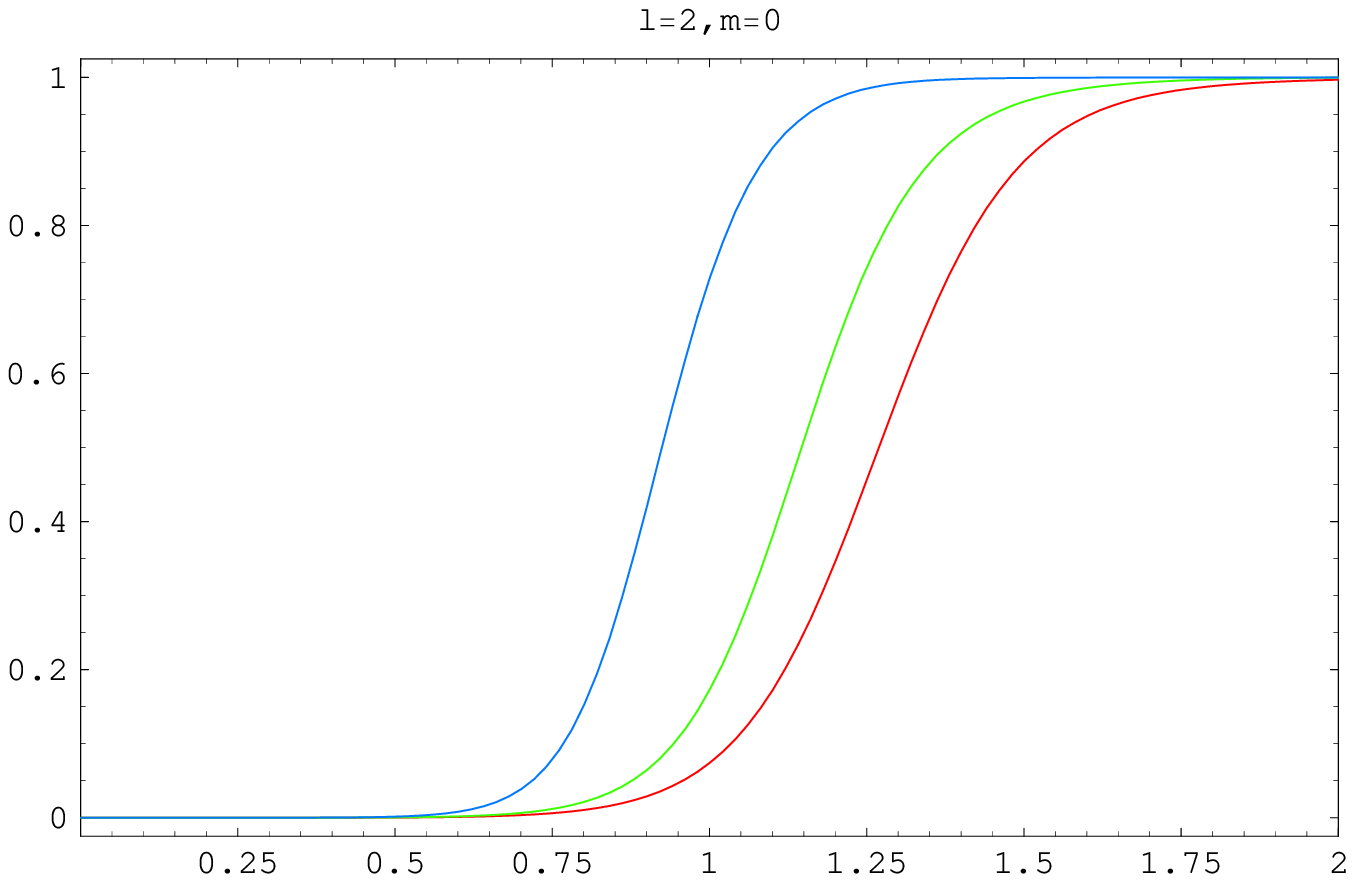}
\end{center}\caption{
Greybody factor for the $D=4+n=5$ black hole
for the brane scalar emission into the $\ell=2$ modes
with $m=-2,2$ for the upper-left, upper-right,
$m=-1,1$ for the middle-left, middle-right,
and $m=0$ for the lower graphs, respectively. The $m>0$ mode
shows the superradiance, namely a negative greybody factor,
in the low-frequency region $\omega<ma/(r_h^2+a^2)$.
The red, green and blue curves correspond to $\astar=0$, $0.5$ and $1.0$.}
\label{greybody_2}
\end{figure}

\subsection{Comparison with analytic expansions for $D=4+n=5$}
The numerical results of the greybody factors for the $D=5$ black
hole is shown in Figs.~\ref{greybody_0}-\ref{greybody_2}.
They are in good agreement with the previous analytic expression
in~\cite{Ida:2002ez} in the region $r_h\omega\lesssim 0.3$.
At $\omega=\omega_0$ with
\begin{align}
  \omega_0 = m\Omega = {ma\over r_h^2+a^2},
  \end{align}
the Bose statics factor diverges, but this divergence is regularized
to give a finite emission rate due to the zero absorption rate
at this point, where greybody factor cross the zero.\footnote{
The expression `Hawking absorption' \cite{Ida:2005zi} to describe
this situation might be misleading.}
As we claimed in Ref.~\cite{Ida:2005zi}, our analytic expression
in Ref.~\cite{Ida:2002ez}
correctly shows for which $\omega$
there emerges the super-radiance with
the negative greybody factor
$\propto \widetilde{Q} = (1+a_*^2)\omega-ma_*<0$.

\begin{figure}[tbp]
\begin{center}
\includegraphics[width=0.4\linewidth]{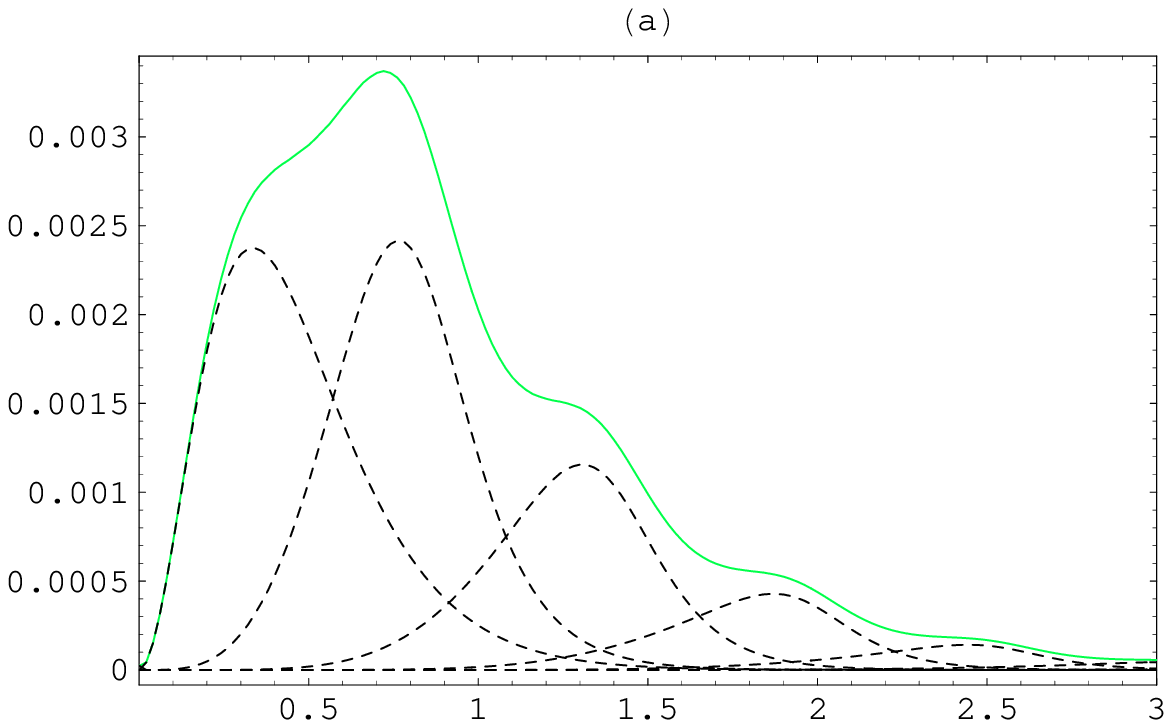}
\includegraphics[width=0.4\linewidth]{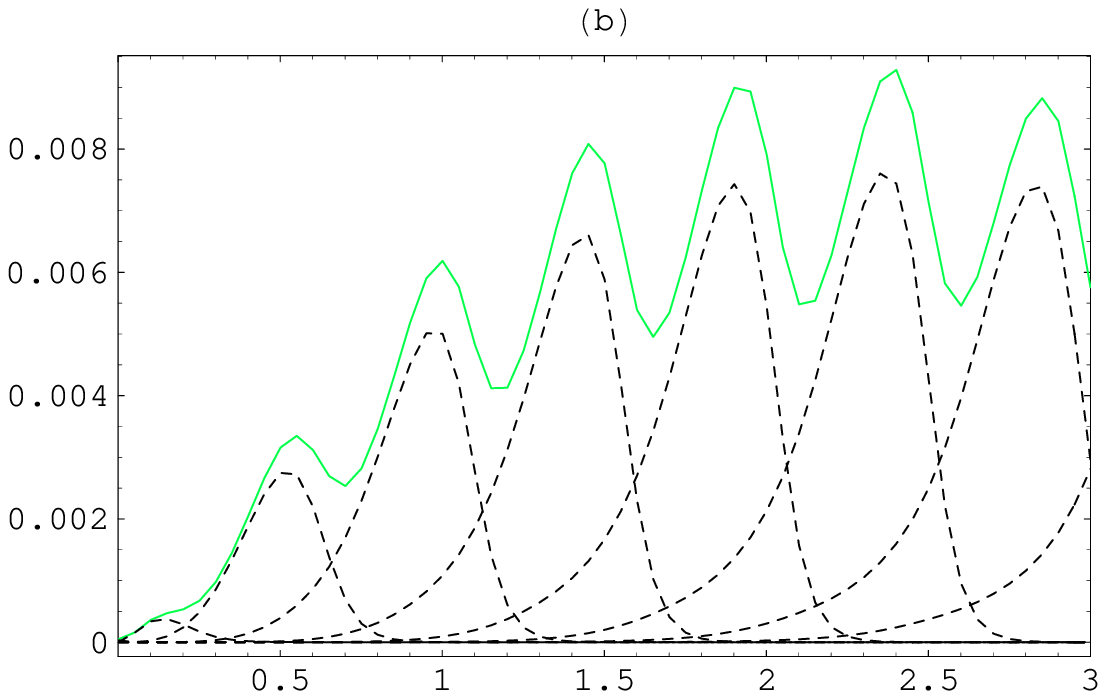}
\end{center}
\caption{The total power spectrum of five dimensional
black hole with $\astar=0.8$ and $1.5$. In Fig.~(a),
The green curve denotes the total power spectrum.
Dotted curves correspond to
$\ell=m=0,1,2,3,4$ and $5$ modes, respectively.
Figure (b) shows that when the hole is rotating
the total power spectrum is
essentially determined by $\ell=m$ modes, with each peak corresponding to
each angular mode (see also Fig.~\ref{elmmodes}).}
\label{powerspectrumtoshowmodes}
\end{figure}
\begin{figure}[tbp]
\begin{center}
\includegraphics[width=0.7\linewidth]{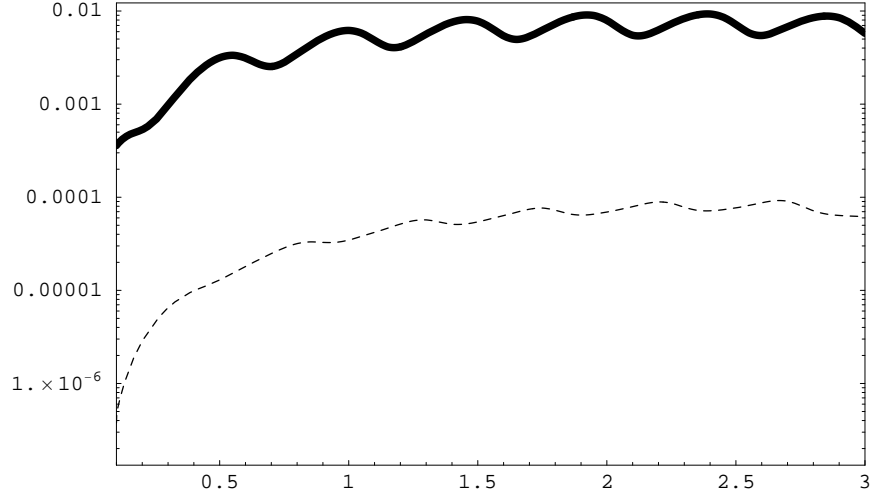}
\end{center}\caption{Total power spectrum of $D=4+n=5$ black hole with $a_*=1.5$.
The small concave curve denote the contribution from
the sum of other than $\ell=m$ modes. We can safely neglect the other
modes than $\ell=m$ modes.}
\label{elmmodes}
\end{figure}
\begin{figure}[tbp]
\begin{center}
\includegraphics[width=0.4\linewidth]{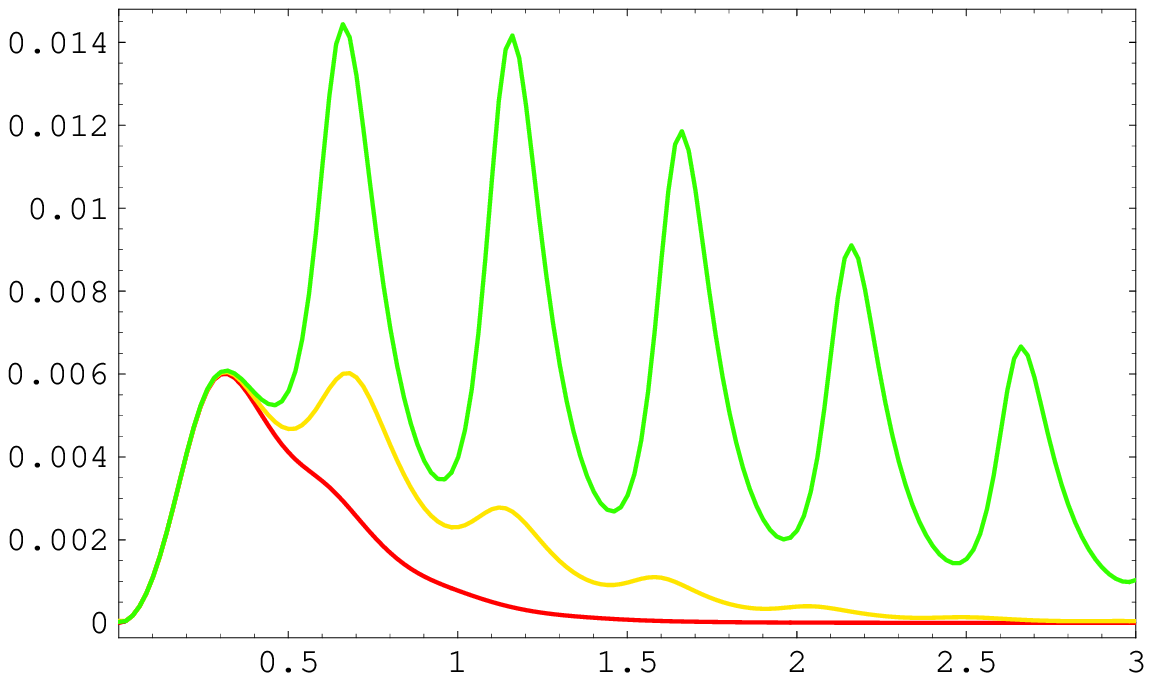}
\includegraphics[width=0.4\linewidth]{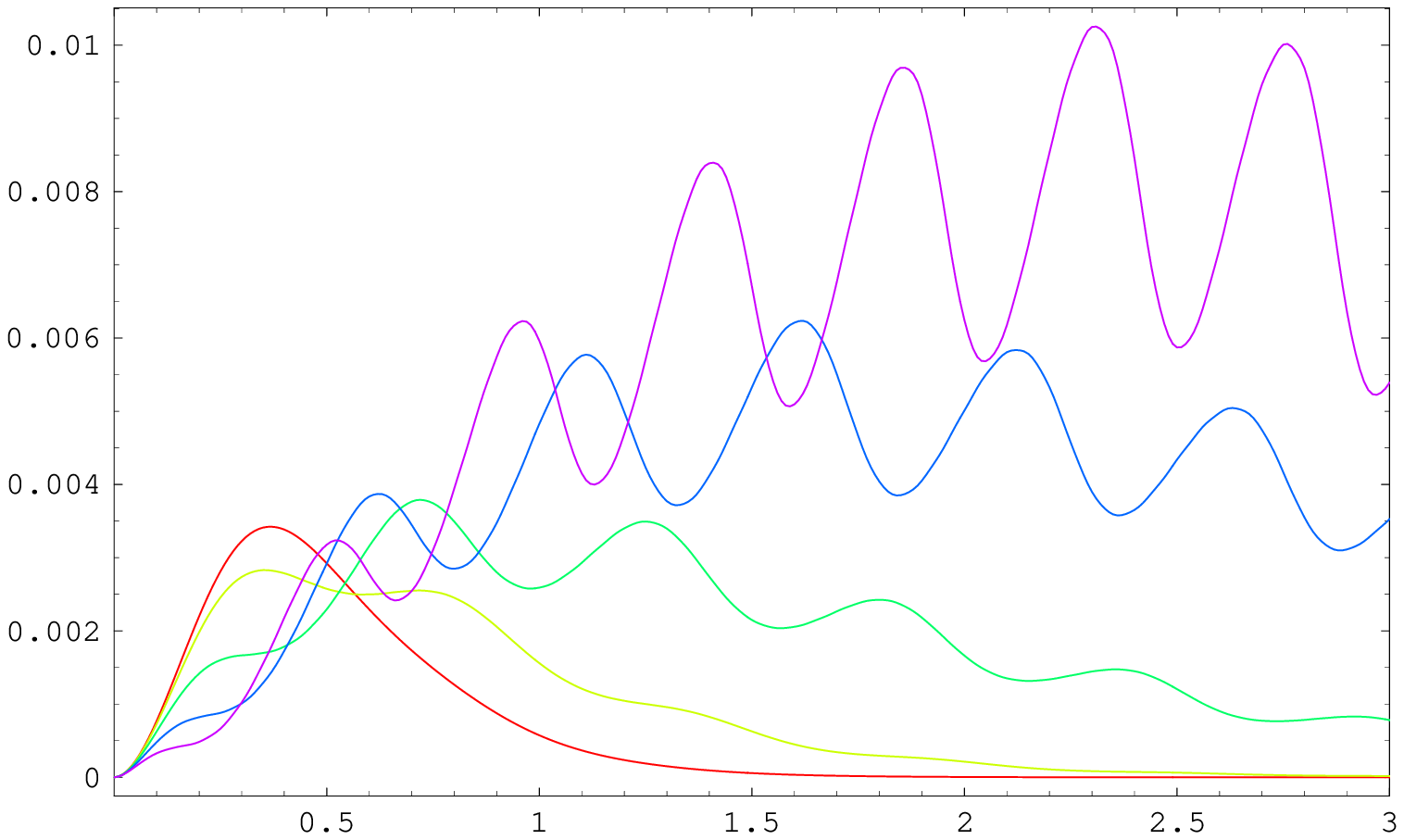}
\includegraphics[width=0.4\linewidth]{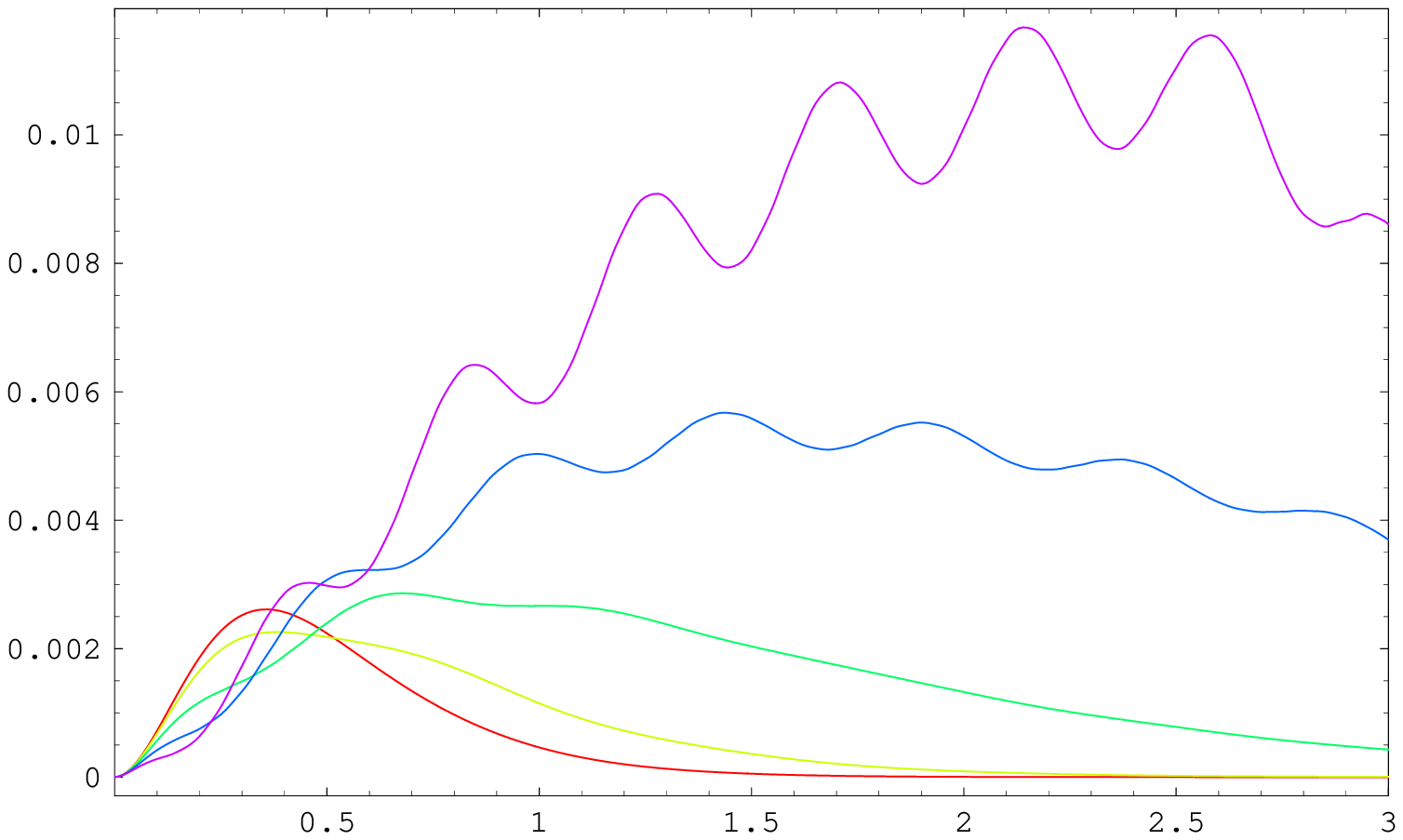}
\includegraphics[width=0.4\linewidth]{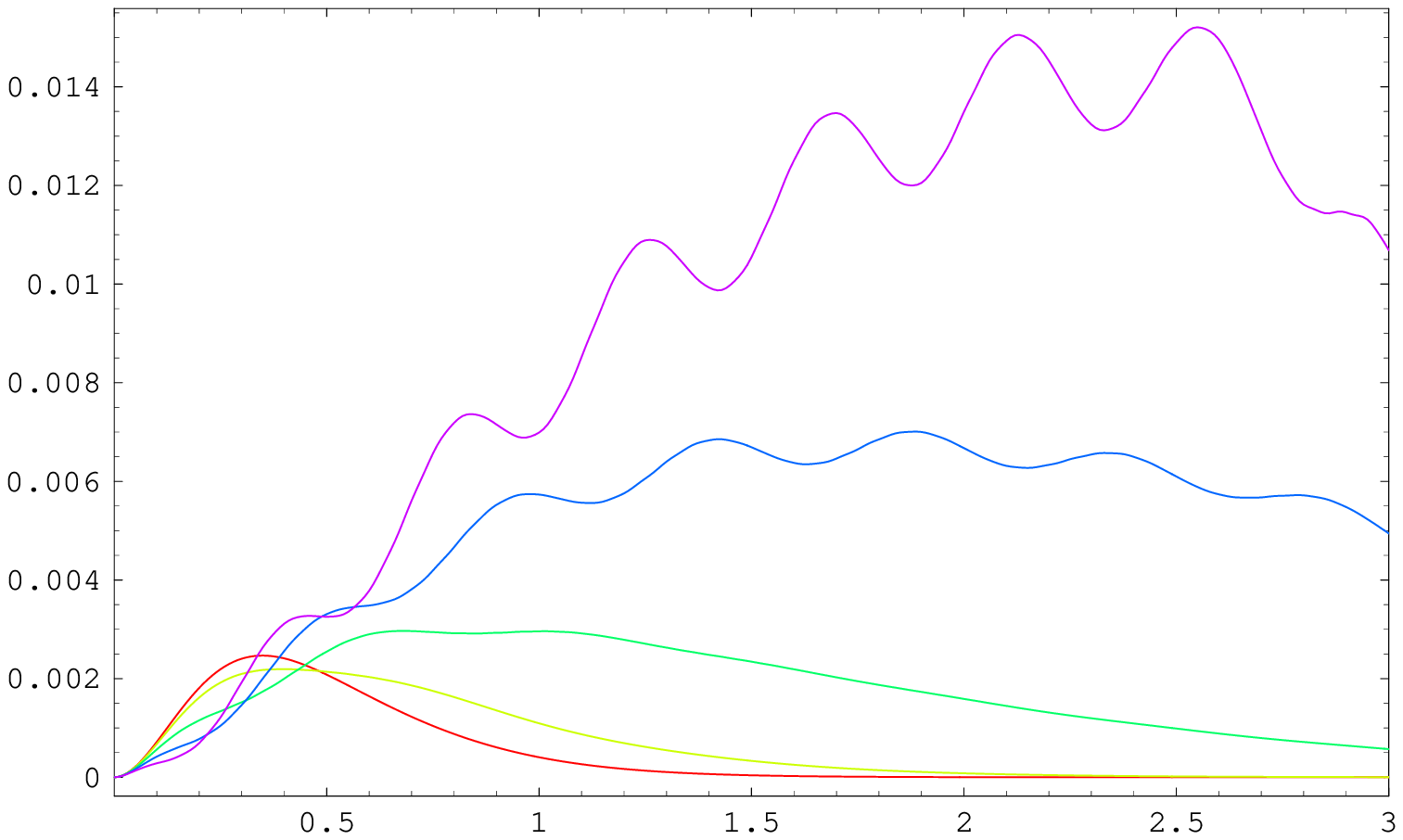}
\end{center}\caption{Power spectrum of a
rotating black hole in $D=4+n=4$, 5, 7, 10
for the upper-left, upper-right, lower-left
and lower-right plots,
respectively. The red,
yellow, green, blue and purple curves denote
$\astar=0$, $0.4$, $0.8$, $1.2$ and $1.6$ respectively.
Each peak corresponds to each angular mode with $\ell=m$,
see Fig.~\ref{powerspectrumtoshowmodes}.}
\label{powerspectrums}
\end{figure}
\begin{figure}[tbp]
\begin{center}
\includegraphics[width=0.4\linewidth]{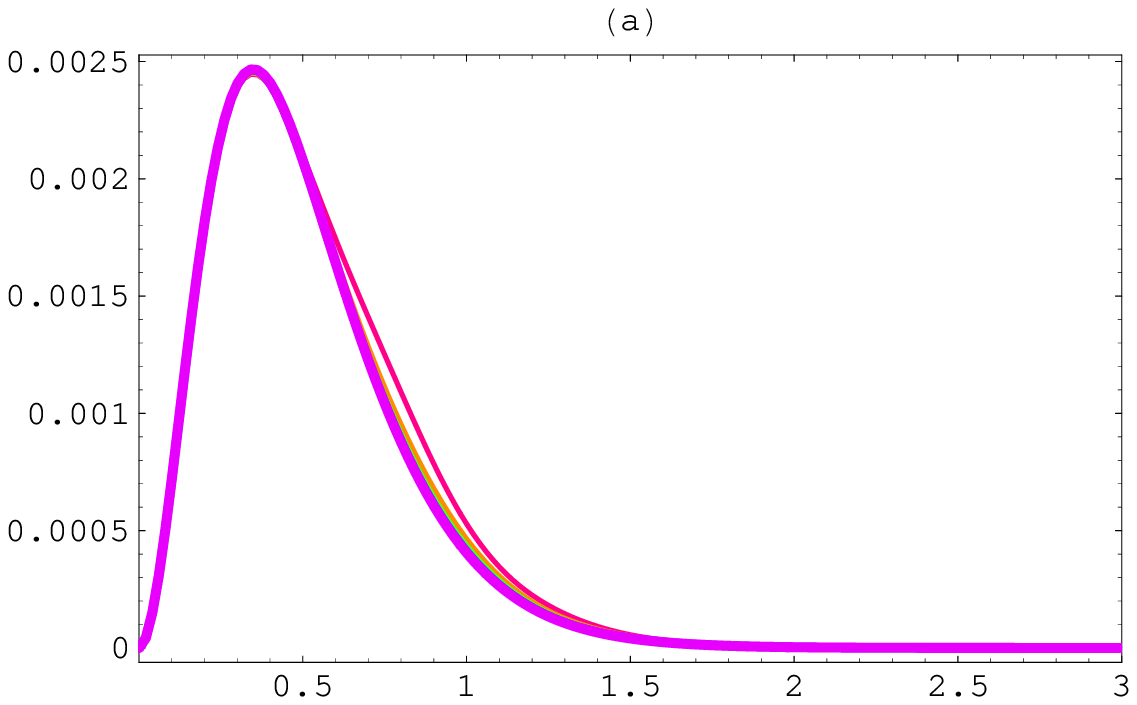}
\includegraphics[width=0.4\linewidth]{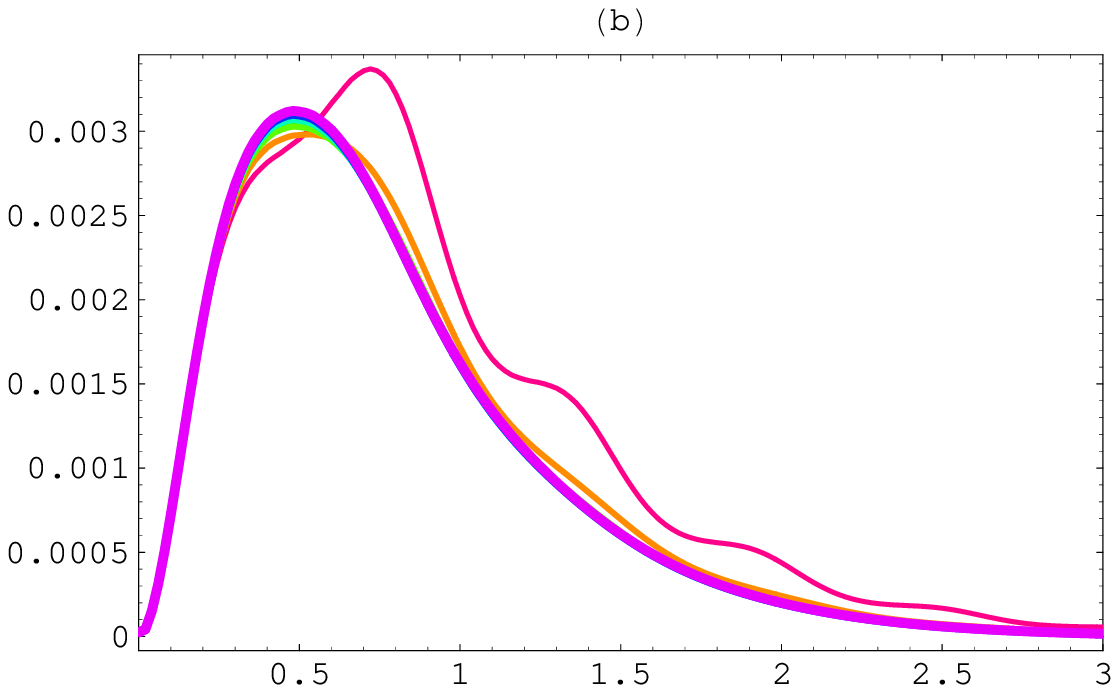}
\includegraphics[width=0.4\linewidth]{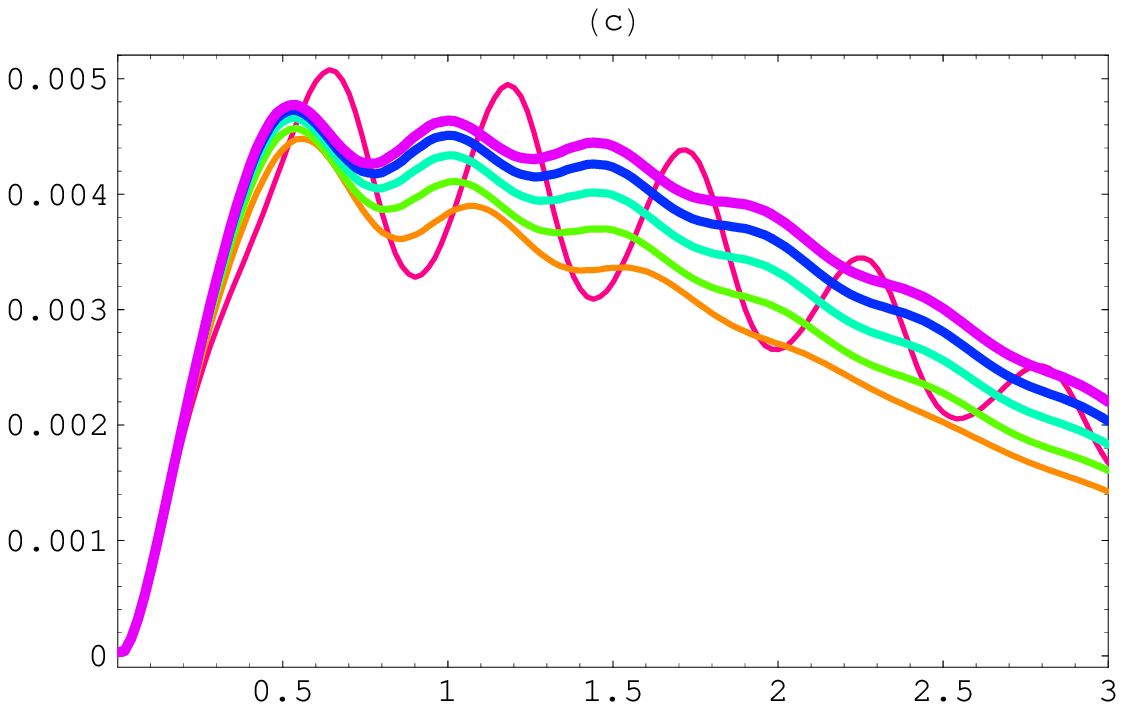}
\includegraphics[width=0.4\linewidth]{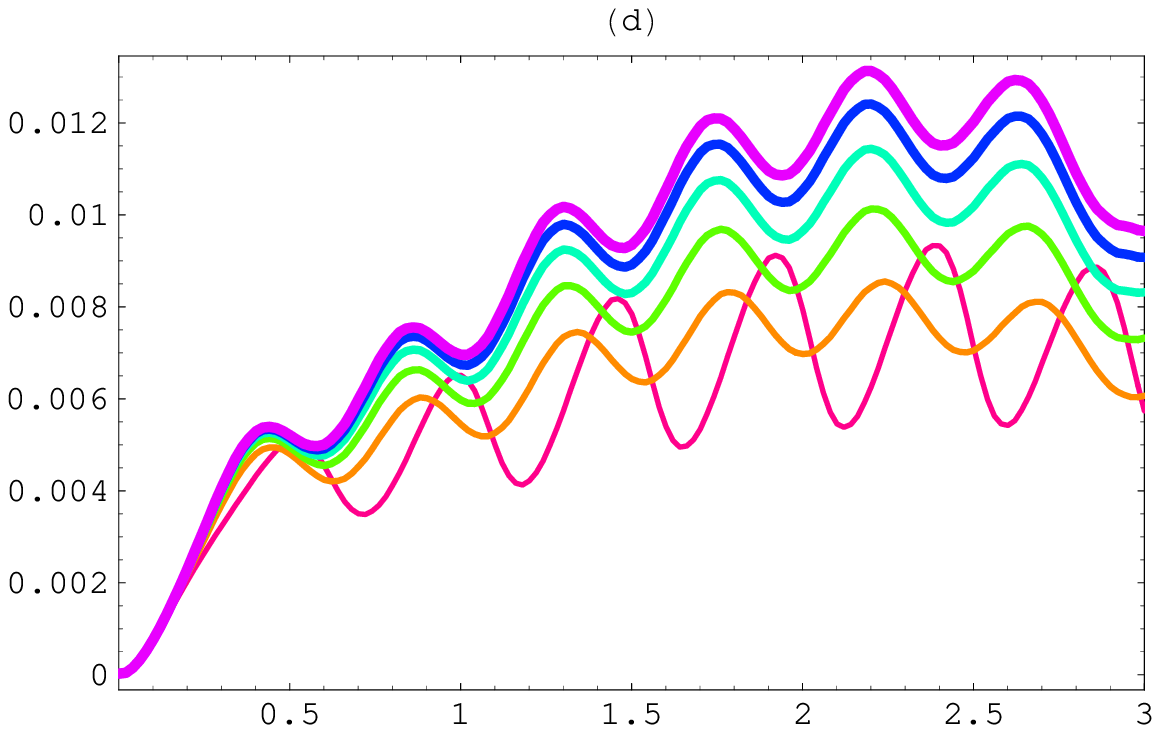}
\end{center}\caption{Total power spectrum of higher dimensional black holes with $\astar=0, 0.5, 1.0$ and $1.5$
for the upper-left, upper-right, lower-left, lower-right plots, respectively.
The pink, red, yellow, green, cyan and blue curves correspond to
$D=4+n=5,6,7,8,9$ and $10$, respectively.}
\label{fullpowerspectrum}
\end{figure}

\subsection{Power spectrum}
In Fig.~\ref{powerspectrumtoshowmodes} we show the power spectrum for $D=4+n=5$ black hole.
This shows that the spectrum is dominated by the $\ell=m$ modes.
The result show that the domination of the $\ell=m$ mode is more significant
for highly rotating case.
As a check we plot the $\ell=m$ contribution and that from
the other modes in Fig.~\ref{elmmodes}.

Fig.~\ref{powerspectrums} we plot using only the $\ell=m$ modes.
We take the angular modes from $\ell=m=0$ to $7$. In the low
energy region ($r_h\omega  \lesssim 0.3$) the spectrum of the
highly rotating black hole is suppressed compared with the
non-rotating power spectrum, confirming the analytic result in low
energy expansions~\cite{Ida:2002ez}, while in the higher energy
regime the spectrum is greatly enhanced. This tendency is stronger
for larger dimensions. To see that more explicitly, we also plot
the power spectrum varying the number of dimensions in
Fig.~\ref{fullpowerspectrum}.

\begin{figure}[tbp]
\begin{center}
\includegraphics[width=0.4\linewidth]{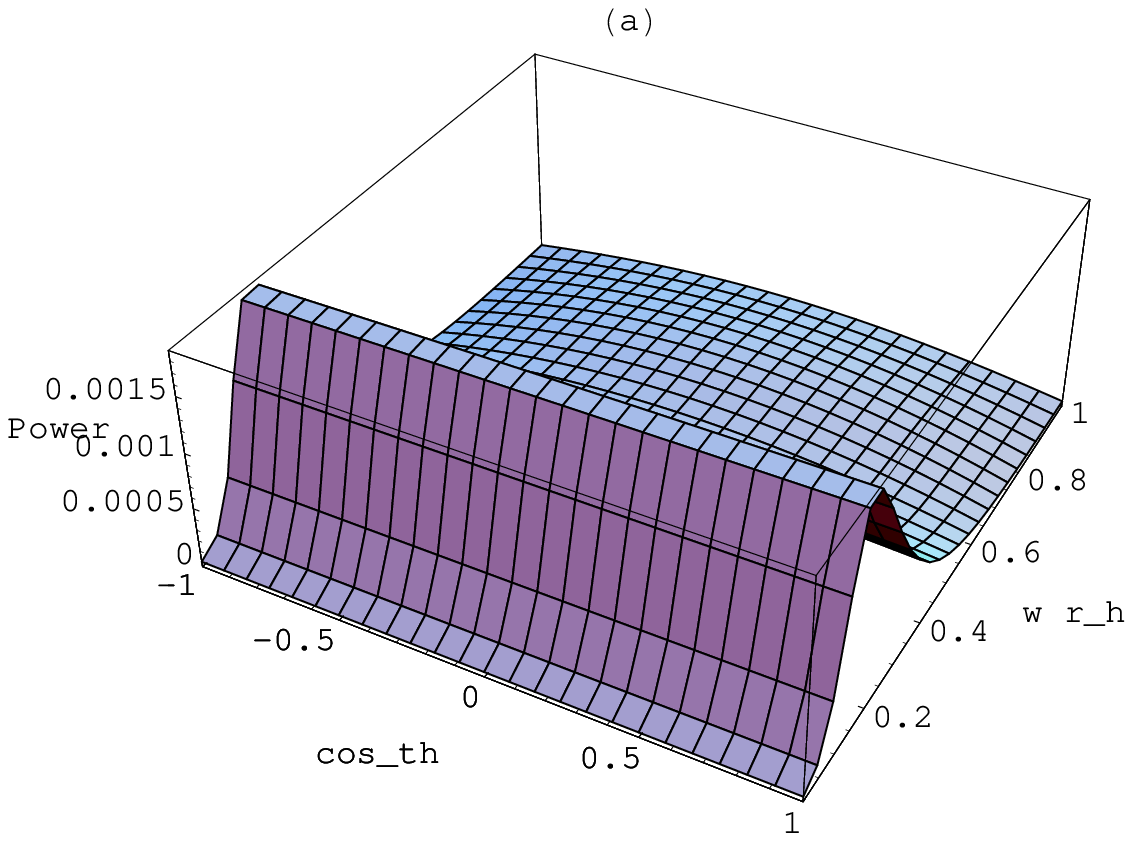}
\includegraphics[width=0.4\linewidth]{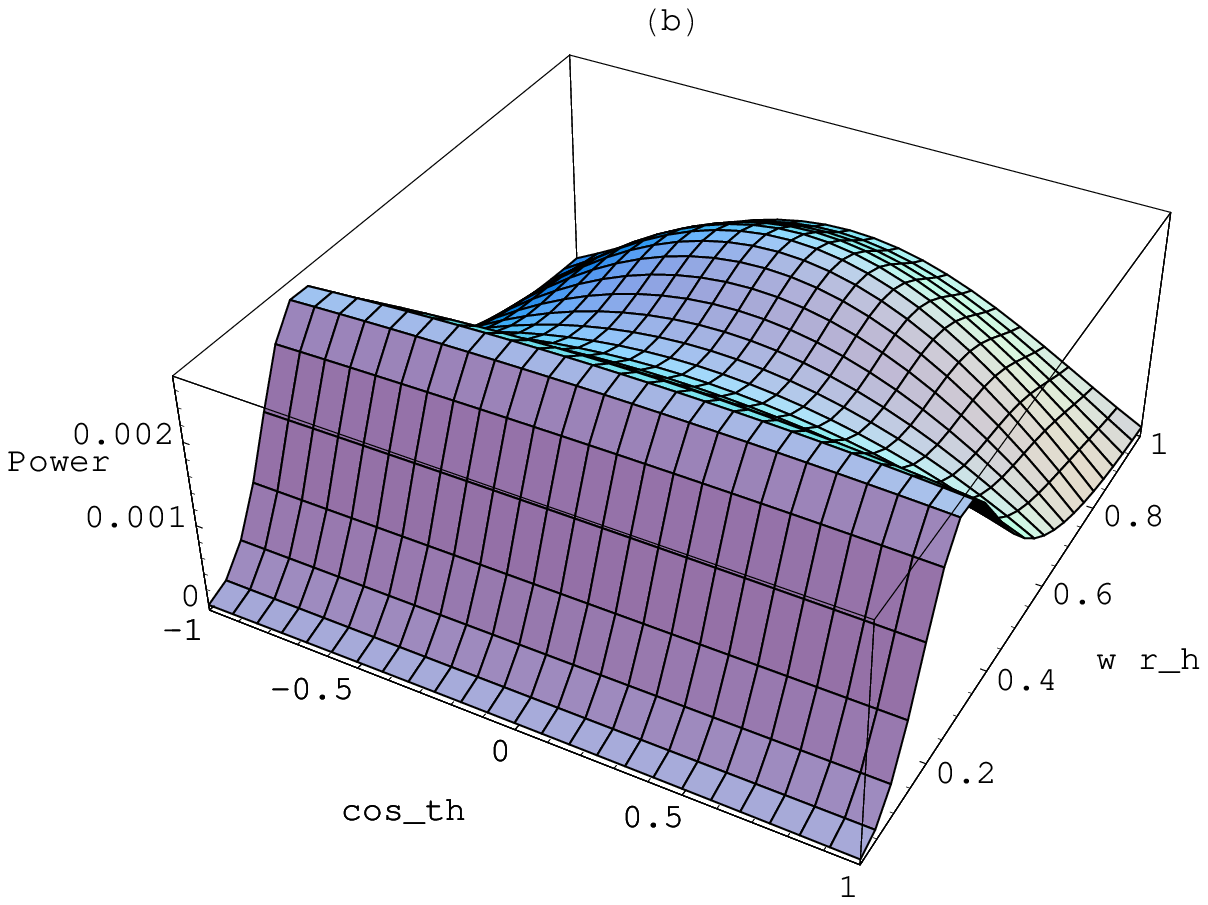}
\includegraphics[width=0.4\linewidth]{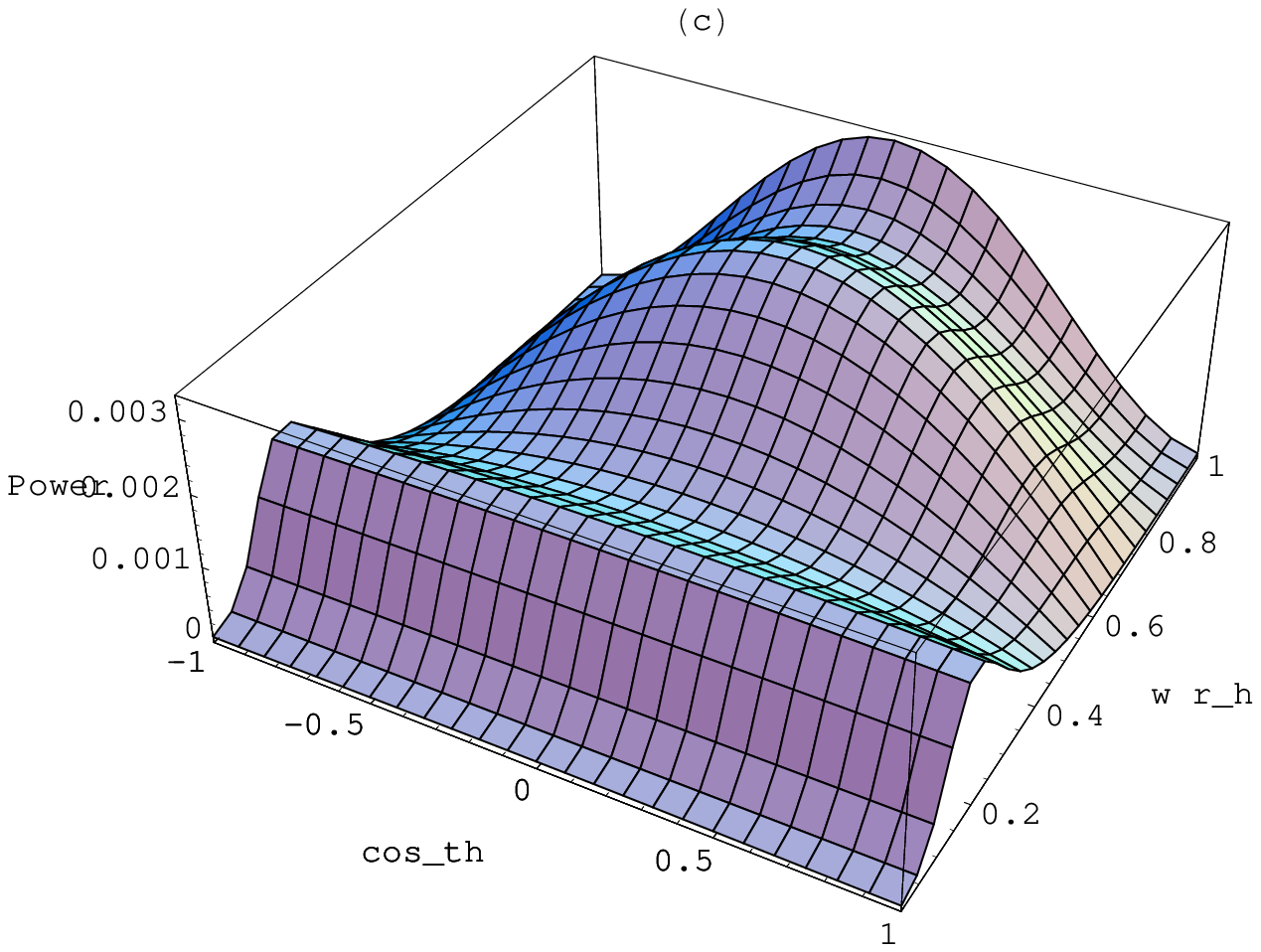}
\includegraphics[width=0.4\linewidth]{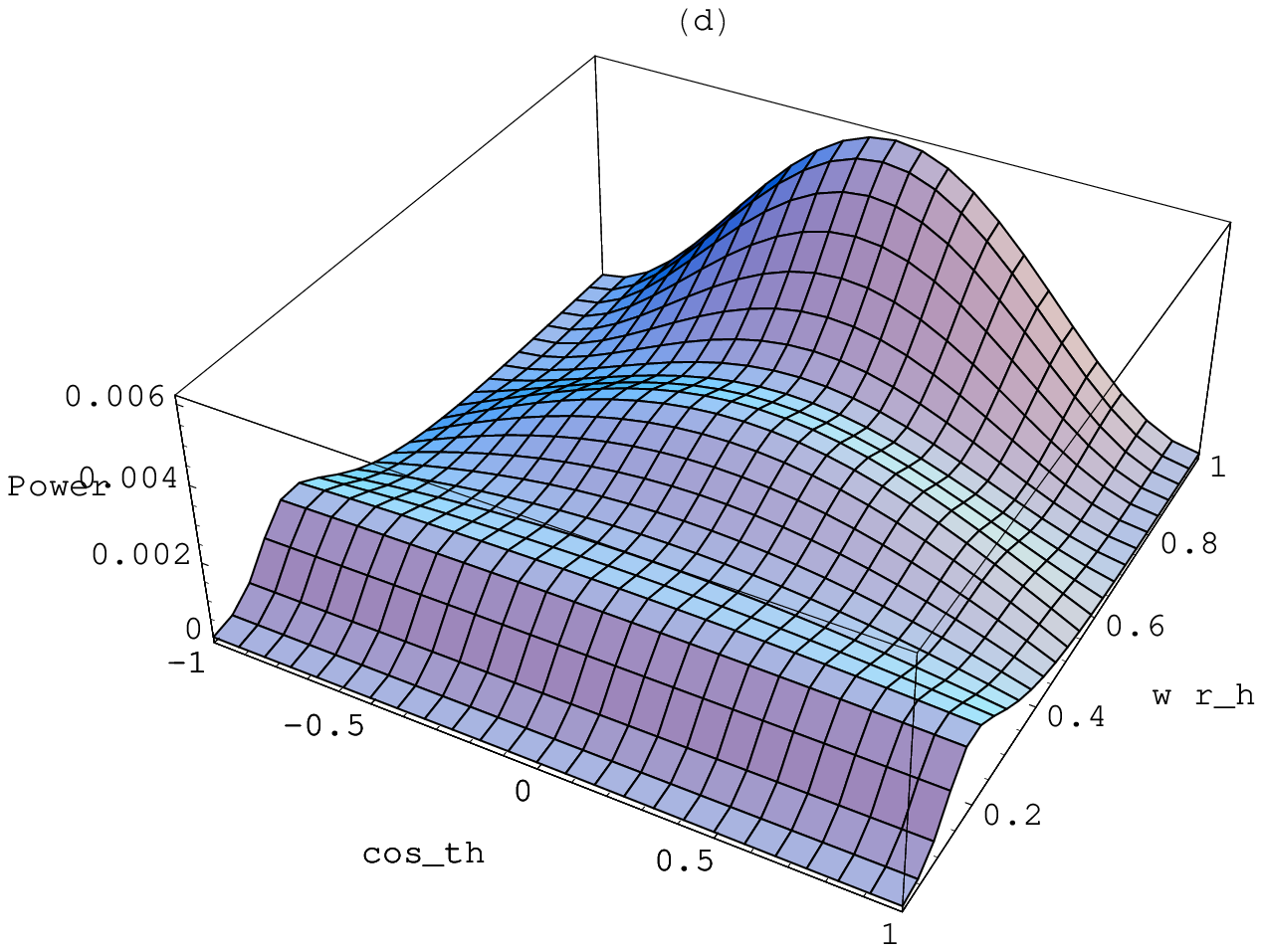}
\end{center}
\caption{Angular distribution for $D=5$ black hole.
The $a_*=0$, $0.5$, $1.0$, $1.5$ from Fig.~(a)  to (d).}
\label{angularpowerspectrum}
\end{figure}

\subsection{Angular distribution}
In Fig.~\ref{angularpowerspectrum} we plot the angular distribution
for $D=5$ black holes. We approximate the spheroidal harmonics
by the spherical harmonics with the assumption $a\omega\ll 1$.
We confirm the previous result in the low frequency approximation~\cite{Ida:2002ez} that the anisotropy is greatly enhanced for a highly rotating black hole, due to the $\ell=m>0$ mode.

\section{Discussion}
We have explained the importance of the angular momentum when
one considers TeV scale black hole production and evaporation.
New numerical results are shown: greybody factor for the brane
scalar emission from a general $D=4+n$ dimensional rotating black
hole without relying on the low frequency expansions.
The greybody factors are obtained for general $D\geq 4$
dimensional cases and the various angular modes ($\ell,m$).
We confirmed the nontrivial angular dependence
of the scalar emission
at the middle energy region $r_h\omega\sim 0.5$
and found that it is even more enhanced at higher
energy region.

To understand the actual evolution of a black hole and to predict
the collider signature, we need further investigations. It is
important to determine the greybody factors for spinor and vector
fields. The evolution of the angular momentum and the mass can be
determined once all the greybody factors are determined
\cite{PageII, TPIII}. In particular, the spin-down phase, whose
time evolution has been impossible to determine so far, can be
precisely described. One can in principle determine the angular
momentum of the produced black hole from the nontrivial angular
distribution of the signals.


\end{document}